\documentstyle[12pt]{article}
\begin{document}

\hfill DUKE-CGTP-98-11

\hfill hep-th/9812084

\vspace{1.0in}

\begin{center}

{\large\bf D-Branes and $\mbox{Spin}^c$ Structures }

\vspace{0.75in}

Robert L. Bryant  \\
Department of Mathematics \\
Box 90320 \\
Duke University \\
Durham, NC  27708 \\
{\tt bryant@math.duke.edu} \\

 $\,$

Eric Sharpe \\
Department of Physics \\
Box 90305 \\
Duke University \\
Durham, NC  27708 \\
{\tt ersharpe@cgtp.duke.edu} \\

 $\,$

\end{center}

It was recently pointed out by E. Witten that for a D-brane
to consistently wrap a submanifold of some manifold, the normal
bundle must admit a $\mbox{Spin}^c$ structure.  We examine this
constraint in the case of type II string compactifications with vanishing
cosmological constant, and argue that in all such cases,
the normal bundle to a supersymmetric cycle is automatically $\mbox{Spin}^c$.

\begin{flushleft}
December 1998 
\end{flushleft}

\newpage

\section{Introduction}

It was recently pointed out by E. Witten \cite{ed1,ed2} that for a D-brane to
consistently wrap a submanifold of some manifold, the normal
bundle must admit a $\mbox{Spin}^c$ structure.

For type II string compactifications of vanishing cosmological constant,
other conditions for a consistently wrapped D-brane were worked
out several years ago \cite{bbs,bsv}.  

In this technical note we shall argue that that for compactifications 
of vanishing cosmological
constant, the $\mbox{Spin}^c$ constraint is redundant,
that the conditions previously known for supersymmetric cycles
always imply 
that the normal bundle admits a $\mbox{Spin}^c$ structure.

What are the options?
In a Calabi-Yau, D-branes can be (supersymmetrically)
wrapped on complex submanifolds
and on special Lagrangian submanifolds \cite{bbs}.
In a $G_2$ holonomy 7-manifold, D-branes can be wrapped on
associative (3-)submanifolds and coassociative (4-)submanifolds \cite{bsv}.
In a $\mbox{Spin}(7)$ 8-manifold, D-branes can be wrapped on
Cayley (4-)submanifolds \cite{bsv}.

In each case, a supersymmetric cycle corresponds to a calibrated
submanifold.  Recall \cite{mcthesis,hl} a calibration is a closed $p$-form 
$\phi$ on
a Riemannian manifold $M$ such that $\phi$ restricts to each tangent
$p$-plane to be less than or equal to the volume form of that $p$-plane.
Oriented submanifolds for which $\phi$ restricts to be equal to the Riemannian
volume form (with respect to the induced metric) are said to be
calibrated by the form $\phi$.  

Calibrated submanifolds have an orientation,
so, as the ambient space is oriented, we know that the structure
group of the normal bundle to a calibrated submanifold is
$SO(n)$, as opposed to $O(n)$.  
Thus, it makes sense to ask whether the normal bundle
admits a $\mbox{Spin}^c$ structure, or even a $\mbox{Spin}$ structure.

We shall begin by making some general observations
concerning $\mbox{Spin}^c$ structures on normal bundles,
then we shall examine each type of calibrated submanifold
on a case-by-case basis.

Throughout this note we shall assume that we have compactified  
on a smooth manifold, and that the D-brane is wrapped on a smooth
submanifold.  In general if these conditions are not satisfied
then there can be nonobvious subtleties.  
Even in complex geometry, if one tries to wrap
a brane on a singular subvariety or on a ``subvariety'' that is
actually a subscheme, one will encounter a number of apparent
difficulties \cite{paulpriv}. 

We shall also assume the reader is acquainted with
the notion of a $\mbox{Spin}^c$ structure.  If not, consult
for example \cite[appendix D]{lm} for expository material.

\section{Generalities}   \label{gen}

Before we begin examining individual cases, we will make
a few comments that will greatly simplify the analysis.

Let $M$ be either a Calabi-Yau, a $G_2$ holonomy 7-manifold,
or a $\mbox{Spin}(7)$ holonomy 8-manifold, and let $P$ be
a calibrated submanifold.
Let $TP$ denote the tangent bundle of $P$, and $N$ its normal
bundle inside $M$.

Bundles with structure group $SU(n)$, $G_2$, and $\mbox{Spin}(7)$
have no characteristic classes\footnote{Because for $G = SU(n)$,
$G_2$, and $\mbox{Spin}(7)$, $\pi_1( BG ) = \pi_2(BG) = 0$.}
living in $H^1(M, {\bf Z}_2)$ or $H^2(M, {\bf Z}_2)$.
Thus, the Stiefel-Whitney classes satisfy 
$w_1(TP \oplus N) = w_2(TP \oplus N) = 0$.
The Whitney sum formula then yields
\begin{eqnarray*}
w_1(N) & = & w_1(TP) \\
w_2(N) & = & w_2(TP) \, + \, w_1(TP)^2
\end{eqnarray*}

Since $P$ is oriented, $w_1(TP) = 0$, and so $w_1(N) = 0$.
Thus, the structure group of the normal bundle $N$ can
be reduced to $SO(n)$, not merely $O(n)$, as remarked in the introduction.
The second relation now implies that $w_2(N) = w_2(TP)$.
It is a standard fact that an $SO(n)$ bundle $N$ will admit
a $\mbox{Spin}^c$ structure if and only if $w_2(N)$ is the mod 2 
reduction of an element of $H^2(P, {\bf Z})$.  Thus, since $w_2(N) =
w_2(TP)$, we see immediately that the normal bundle $N$ will admit
a $\mbox{Spin}^c$ structure if and only if the tangent bundle
to the calibrated submanifold admits a $\mbox{Spin}^c$ structure.
(For that matter, $N$ will admit a $\mbox{Spin}$ structure if and only
if $TP$ admits a $\mbox{Spin}$ structure.)

We shall see, on a case-by-case basis, that for associative and
coassociative submanifolds
of $G_2$ holonomy 7-manifolds, Cayley submanifolds of $\mbox{Spin}(7)$
holonomy 8-manifolds, complex submanifolds of Calabi-Yau's, and
special Lagrangian submanifolds of Calabi-Yau's of complex dimension
less than five, the tangent bundle to the calibrated submanifold
admits a $\mbox{Spin}^c$ structure, and so the normal bundle
necessarily admits a $\mbox{Spin}^c$ structure.

In addition, in each case we will also try to give additional
information regarding normal bundles.  For example, in several
cases normal bundles are necessarily trivializable, not just
$\mbox{Spin}^c$.  We shall also give relevant technical pointers
regarding $\mbox{Spin}^c$ structures in general.

\section{Calabi-Yau manifolds}

\subsection{Complex submanifolds}

Complex submanifolds of a Calabi-Yau can be viewed as calibrated
submanifolds \cite[section 0]{mcthesis}; 
simply choose $\phi = \frac{1}{p!} \omega^p$
for a complex $p$-fold submanifold, where $\omega$ is the K\"ahler
form on the Calabi-Yau.  

As mentioned in section~(\ref{gen}), the normal bundle to a 
calibrated submanifold will admit a $\mbox{Spin}^c$ structure
if and only if the tangent bundle to the calibrated submanifold
admits a $\mbox{Spin}^c$ structure.  Now, it is a standard fact
that the tangent bundle of any complex manifold (in fact,
any almost complex manifold) admits a $\mbox{Spin}^c$ structure
\cite[appendix D]{lm}, thus the normal bundle to a
complex submanifold must admit a $\mbox{Spin}^c$ structure.

There is a more direct way to get this result.
The normal bundle to a complex submanifold of a Calabi-Yau
is a $U(n)$ bundle, and so always admits a $\mbox{Spin}^c$ structure
\cite[appendix D]{lm}.

\subsection{Special Lagrangian submanifolds}

Special Lagrangian submanifolds are calibrated submanifolds
defined by a calibration $\phi$ equal to the real part of the
holomorphic $n$-form
trivializing the canonical bundle of the Calabi-Yau.
(It can be shown \cite[section III.1]{hl} that a submanifold
is special Lagrangian if and only if it is Lagrangian and 
the restriction of the imaginary part of the holomorphic $n$-form
vanishes.)

In general one does not expect the special Lagrangian submanifolds
of any Calabi-Yau to have $\mbox{Spin}^c$ normal bundles, but
in the special case of low-dimensional Calabi-Yau's (in particular, Calabi-Yau's
of complex dimension less than five) we shall see that the normal
bundle to any special Lagrangian submanifold is $\mbox{Spin}^c$.

First, note that this is trivial to check for Calabi-Yau's of complex dimension
one or two, so we shall only consider Calabi-Yau's of dimensions three
and four.

For Calabi-Yau's of complex dimension three, the special Lagrangian
submanifold will be a compact oriented 3-manifold.
It is known that the tangent bundle of any compact oriented 3-manifold
is trivializable \cite[problem 12-B]{ms}, 
thus it admits a $\mbox{Spin}^c$ structure and so by the arguments
of section~(\ref{gen}), the normal bundle must also admit
a $\mbox{Spin}^c$ structure.

In fact, in this case we can make a much stronger statement.
It can be shown \cite[corollary 3-3]{mcthesis} that the normal
bundle of a special Lagrangian submanifold is isomorphic to the
tangent bundle of the special Lagrangian submanifold.
Thus, since the tangent bundle of a compact oriented 3-manifold
is trivial, we know that the normal bundle to a special Lagrangian
submanifold of a Calabi-Yau 3-fold is trivializable, not just
$\mbox{Spin}^c$.

For Calabi-Yau's of complex dimension four, we use the standard
fact that the tangent bundle of any oriented compact four-manifold
admits a $\mbox{Spin}^c$ structure \cite[lemma 3.1.2]{morgan}.
Thus, by either the arguments of section~(\ref{gen}) or 
since the normal bundle of a special Lagrangian submanifold is
isomorphic to the tangent bundle, we see that the
normal bundle to a special Lagrangian submanifold necessarily
admits a $\mbox{Spin}^c$ structure.

For Calabi-Yau's of higher dimension, it is not clear that in general
normal bundles to their special Lagrangian submanifolds will
admit $\mbox{Spin}^c$ structures.  However, as higher dimensional
Calabi-Yau's cannot be used in type II string compactifications,
this is not a relevant issue.

\section{$G_2$ 7-manifolds}

On a 7-manifold $M$ of $G_2$ holonomy, there exists a 3-form $\phi$
that is compatible with the $G_2$ structure \cite{joyceg21}.  For example,
on ${\bf R}^7$ let $y_1, y_2, \cdots y_7$ denote an oriented, orthonormal
basis, then define 
\begin{displaymath}
\phi \: = \: y_1 \wedge y_2 \wedge y_7 \: + \: y_1 \wedge y_3 \wedge y_6
\: + \: y_1 \wedge y_4 \wedge y_5 \: + \: y_2 \wedge y_3 \wedge y_5
\: - \: y_2 \wedge y_4 \wedge y_6 \: + \: y_3 \wedge y_4 \wedge y_7
\: + \: y_5 \wedge y_6 \wedge y_7
\end{displaymath}
Then the subgroup of $GL_+(7, {\bf R})$ (the orientation-preserving
subgroup of $GL(7,{\bf R})$) that preserves $\phi$ is precisely $G_2$.
The Hodge dual to $\phi$, namely $* \phi$, has analogous properties.

On a $G_2$ holonomy 7-manifold, there are two natural sets of calibrated
submanifolds \cite{joyceg22}.  One set is defined by taking the calibration to
be $\phi$, and consists of three-dimensional submanifolds known as
associative submanifolds.  The other set is defined by taking
the calibration to be $* \phi$, and consists of four-dimensional
submanifolds known as coassociative submanifolds.
We shall study normal bundles to each type of calibrated submanifold
separately.

\subsection{Associative (3-)submanifolds}   \label{g2assoc}

We shall argue that the normal bundle to an associative
submanifold is not just $\mbox{Spin}^c$, but actually
trivializable.

First, since the tangent bundle of an oriented compact 3-manifold
is trivial \cite[problem 12-B]{ms}
and so trivially $\mbox{Spin}^c$, we know from the
arguments given in section~(\ref{gen}) that the normal bundle
to an associative submanifold admits a $\mbox{Spin}^c$ structure.

In fact, we can make a much stronger statement concerning the
normal bundle to an associative submanifold, namely that it is
trivializable.
We shall need a few general facts concerning associative submanifolds
\cite[section 5]{mcthesis}.  First, the restriction of the tangent
bundle of a $G_2$ holonomy 7-manifold to an associative submanifold
is a principal $SO(4) = Sp(1) \times_{{\bf Z}_2} Sp(1)$ bundle.
Write ${\bf R}^7 = {\bf H} \oplus \mbox{im } {\bf H}$, where
${\bf H}$ denotes the quaternions.  (Intuitively, if we locally
identify tangent directions to the $G_2$ holonomy 7-manifold with
vectors in ${\bf R}^7$, then
directions in $\mbox{im }{\bf H}$ will be tangent to the associative
submanifold, and directions in ${\bf H}$ will be normal to the
associative submanifold.)  Then the structure group of the
restriction of the tangent bundle of the $G_2$ holonomy 7-manifold
to an associative submanifold can be written as a subgroup of
$GL({\bf H}) \times GL(\mbox{im }{\bf H}) \subset GL(7,{\bf R})$ of
the form 
\begin{displaymath}
\left[ \begin{array}{cc}
       R_q L_p & 0 \\
       0 & \rho(R_q) 
       \end{array} \right]
\end{displaymath}
where $p,q \in {\bf H}$, where $R_q$ and $L_q$
denote right and left multiplication, respectively,
by a quaternion $q \in {\bf H}$,
and where $\rho$ is the irreducible representation of the action of $Sp(1)$ in
$\mbox{im }{\bf H}$ (defined by $\rho(R_q) = L_{\overline{q}} R_q$).  The lower right block, a subset of $GL(\mbox{im }
{\bf H})$, corresponds to the structure group of the tangent bundle
to the associative submanifold, and the upper left block,
a subset of $GL({\bf H})$, corresponds to the structure group of the
normal bundle to the associative submanifold.

Now, the tangent bundle of a compact oriented 3-manifold is topologically
trivial \cite[problem 12-B]{ms}, so the action of $Sp(1)$ given in the
form above by $R_q$ can be gauged away.  Thus, the structure group
of the normal bundle to an associative submanifold is actually
$Sp(1) = SU(2)$.

Finally, $SU(2)$ bundles on a 3-manifold are topologically
trivial\footnote{This is essentially because
$\pi_1( BSU(2) = {\bf H} {\bf P}^{\infty}) = \pi_2({\bf H}{\bf P}^{\infty})
= \pi_3( {\bf H}{\bf P}^{\infty}) = 0$.}, so we see that
the normal bundle to an associative (3-)submanifold is topologically
trivial.  Thus, it trivially admits a $\mbox{Spin}^c$ structure,
and even a $\mbox{Spin}$ structure.

\subsection{Coassociative (4-)submanifolds}

A standard fact concerning 4-manifolds is that the tangent bundle
to any compact oriented 4-manifold admits a $\mbox{Spin}^c$ structure
\cite[lemma 3.1.2]{morgan}, so by the arguments given in
section~(\ref{gen}), we know that the normal bundle
to a coassociative submanifold is $\mbox{Spin}^c$.

As a check, we shall derive this result an alternative way.
It can be shown \cite[proposition 4-2]{mcthesis} that the normal bundle
to a coassociative submanifold $M$ is isomorphic to the bundle of
anti-self-dual two-forms, $\Lambda^2_- T^* M$.  Also, as mentioned
above the tangent bundle to any compact oriented 4-manifold
admits a $\mbox{Spin}^c$ structure.

We claim these two facts imply that $\Lambda^2_- T^* M$ admits a 
$\mbox{Spin}^c$ structure.  It is somewhat easier to explain
why if $M$ were $\mbox{Spin}$ then $\Lambda^2_- T^* M$ would be $\mbox{Spin}$,
so we shall do this first.  If $M$ were $\mbox{Spin}$, 
the the structure group of $TM$ (and more relevantly, $T^*M$)
could be lifted from $SO(4)$ to $\mbox{Spin}(4) = SU(2) \times SU(2)$.
Now, the structure groups of $\Lambda^2_- T^*M$ and $\Lambda^2_+ T^*M$
are in general $SO(3)$ groups associated to the two $SU(2)$ factors
of the structure group of $T^*M$, via the decomposition 
$SO(4) = [ SU(2) \times SU(2) ] / {\bf Z}_2$.  If the structure
group of $T^*M$ can be lifted to $\mbox{Spin}(4) = SU(2) \times SU(2)$,
then clearly the structure groups of $\Lambda^2_+ T^*M$ and $\Lambda^2_-
T^*M$ can both be lifted from $SO(3)$ to $\mbox{Spin}(3) = SU(2)$.

Now, in the case at hand, $M$ is $\mbox{Spin}^c$, but not necessarily
$\mbox{Spin}$.  An argument similar to the one above 
shows that $\Lambda^2_+ T^*M$ and
$\Lambda^2_- T^*M$ are necessarily $\mbox{Spin}^c$ also.
The point is simply that if the structure group of $TM$,
and more to the point $T^*M$, can be lifted from $SO(4)$ to
$\mbox{Spin}^c(4) = SO(4) \times_{{\bf Z}_2} U(1)$,
then the projection into the self-dual and anti-self-dual pieces
of $SO(4)$ yields a lift from $SO(3)$ to $\mbox{Spin}^c(3) = SO(3)
\times_{{\bf Z}_2} U(1)$, consequently both $\Lambda^2_+ T^*M$
and $\Lambda^2_- T^*M$ are $\mbox{Spin}^c$.

Thus, since any compact oriented 4-manifold is $\mbox{Spin}^c$,
we know that $\Lambda^2_+ T^*M$ is $\mbox{Spin}^c$, 
hence the normal bundle to a coassociative (4-)submanifold
is necessarily $\mbox{Spin}^c$.

\section{Cayley (4-)submanifolds of $\mbox{Spin}(7)$ 8-manifolds}

On a $\mbox{Spin}(7)$ holonomy 8-manifold, there is a natural
4-form \cite{joycesp}.  This 4-form can be used as a calibration,
defining real four-dimensional submanifolds known as Cayley
submanifolds.

As the tangent bundle to any compact oriented 4-manifold
necessarily admits a $\mbox{Spin}^c$ structure, 
we know from the arguments of section~(\ref{gen}) that 
the normal bundle to a Cayley submanifold admits a $\mbox{Spin}^c$
structure.

We can also study normal bundles to Cayley submanifolds directly.
The restriction of the tangent bundle of a $\mbox{Spin}(7)$ holonomy
8-manifold to a Cayley submanifold is associated to a principal
$[ Sp(1) \times Sp(1) \times Sp(1) ] / {\bf Z}_2$ bundle
\cite[section 6]{mcthesis}.  Write ${\bf R}^8 = {\bf H} \oplus
{\bf H}$, where $H$ denotes the quaternions.  (Intuitively,
if we locally identify tangent directions to the $\mbox{Spin}(7)$
holonomy 8-manifold with vectors in ${\bf R}^8$, then directions
in one copy of ${\bf H}$ will be tangent to the Cayley submanifold
and directions in the other ${\bf H}$ will be normal to the Cayley
submanifold.)  Then the structure group of the restriction of the tangent
bundle of the $\mbox{Spin}(7)$ holonomy 8-manifold to a Cayley
submanifold can be written as a subgroup of $GL({\bf H}) \times
GL({\bf H}) \subset GL(8,{\bf R})$ of the form
\begin{displaymath}
\left[ \begin{array}{cc}
        L_q R_{p_1} & 0 \\
        0 & L_q R_{p_2} 
        \end{array}  \right]
\end{displaymath}
where $p_1, p_2, q \in {\bf H}$, and other notation is as in
section~(\ref{g2assoc}).  One block corresponds to the structure
group of the tangent bundle of the Cayley submanifold; the other
block corresponds to the structure group of the normal bundle.
It should be clear that the normal bundle is $\mbox{Spin}$
or $\mbox{Spin}^c$ if and only if the tangent bundle to the Cayley
submanifold is $\mbox{Spin}$ or $\mbox{Spin}^c$, respectively.

In passing we shall also mention that
if $X$ is an 
oriented simply-connected 4-manifold,
then all elements of $H^2(X, {\bf Z}_2)$ are mod 2 reductions of
elements of $H^2(X, {\bf Z})$ \cite[section 1.1, p. 6]{dk}.
The condition for an $SO(n)$ bundle to admit a $\mbox{Spin}^c$
structure is that its second Stiefel-Whitney class $w_2 \in
H^2(X, {\bf Z}_2)$ be the mod 2 reduction of an element of
$H^2(X, {\bf Z})$,
consequently all $SO(n)$ bundles
on $X$ admit $\mbox{Spin}^c$ structures.
Thus, if a compact Cayley submanifold is simply-connected,
then one can see immediately that
its normal bundle necessarily admits a $\mbox{Spin}^c$ structure.

\section{Conclusions}

In this technical note we have argued that the normal bundles
to all calibrated submanifolds encountered in 
(supersymmetric) wrapped branes in type II string compactifications
with vanishing cosmological constant admit $\mbox{Spin}^c$ structures.
More precisely, all associative and coassociative submanifolds
of $G_2$ holonomy 7-manifolds, all complex submanifolds of
Calabi-Yau's, all special Lagrangian submanifolds of Calabi-Yau's
of complex dimension less than five, and all
Cayley submanifolds of $\mbox{Spin}(7)$ 8-manifolds have
normal bundles admitting a $\mbox{Spin}^c$ structure.

Thus, for type II string compactifications with vanishing cosmological
constant, the recent observation by E. Witten \cite{ed1,ed2}
that normal bundles must be $\mbox{Spin}^c$ for consistent wrapped
branes, is redundant.  For wrapped branes in $AdS$ compactifications,
by contrast, normal bundles are not automatically $\mbox{Spin}^c$,
and so in that case this constraint is much more interesting. 

In retrospect this is not very surprising.  First, as a rule
of thumb it is relatively
easy to satisfy the $\mbox{Spin}^c$ constraint.  Second, for several
years now various authors have studied wrapped branes in theories
with vanishing cosmological constant without running into any
unexpected anomalies.  Thus, one should not be too surprised
that for compactifications with vanishing cosmological constant,
the $\mbox{Spin}^c$ condition is always satisfied 
automatically.  For wrapped branes in $AdS$ compactifications,
the $\mbox{Spin}^c$ constraint is doubtless more interesting.

\section{Acknowledgements}

E.S. would like to thank P. Aspinwall, A. Knutson,
D. Morrison, and R. Plesser for useful conversations.


\begin{thebibliography}{299}

\bibitem{ed1} E. Witten, ``Baryons and branes in anti-de-Sitter
space'', JHEP {\bf 07} (1998) 6, {\tt hep-th/9805112}. 

\bibitem{ed2} E. Witten, ``D-branes and $K$ theory,'' {\tt hep-th/9810188}.

\bibitem{bbs} K. Becker, M. Becker, and A. Strominger, ``Fivebranes,
membranes, and non-perturbative string theory,'' Nucl. Phys. {\bf B456}
(1995) 130, 
{\tt hep-th/9507158}.

\bibitem{bsv} M. Bershadsky, C. Vafa, and V. Sadov, ``D-branes and
topological field theories,'' Nucl. Phys. {\bf B463} (1996) 420,
{\tt hep-th/9511222}.

\bibitem{mcthesis} R. C. McLean, {\it Deformations and Moduli of Calibrated
Submanifolds}, Ph.D. thesis, Duke University, 1990.

\bibitem{hl} R. Harvey and H. B. Lawson Jr., ``Calibrated geometries,''
Acta Math. {\bf 148} (1982) 47.

\bibitem{paulpriv} P. Aspinwall, S. Katz, and D. Morrison,
in preparation.

\bibitem{lm} H. B. Lawson Jr. and M.-L. Michelsohn,
{\it Spin Geometry}, Princeton University Press, Princeton, 1989.

\bibitem{ms} J. Milnor and J. Stasheff, {\it Characteristic Classes},
Princeton University Press, Princeton, 1974.

\bibitem{morgan} J. W. Morgan, {\it The Seiberg-Witten Equations and
Applications to the Topology of Smooth Four-Manifolds}, Princeton
University Press, Princeton, 1996.

\bibitem{joyceg21} D. Joyce, ``Compact riemannian 7-manifolds with
holonomy $G_2$, I,'' J. Diff. Geom. {\bf 43} (1996) 291.

\bibitem{joyceg22} D. Joyce, ``Compact riemannian 7-manifolds with
holonomy $G_2$, II,'' J. Diff. Geom. {\bf 43} (1996) 329.

\bibitem{joycesp} D. Joyce, ``Compact 8-manifolds with holonomy
$\mbox{Spin}(7)$,'' Invent. Math. {\bf 123} (1996) 507.

\bibitem{dk} S. K. Donaldson and P. B. Kronheimer, {\it The Geometry
of Four-Manifolds}, Oxford University Press, Oxford, 1990.



\end{thebibliography}
\end{document}